\documentclass[a4paper,english,showpacs,floatfix,11pt]{revtex4}
\usepackage[dvips]{graphicx}
\usepackage{longtable}
\usepackage{amsmath,amssymb,amsfonts,amscd}
\usepackage{epsfig}
\usepackage{dcolumn}
\usepackage{latexsym}
\usepackage{babel}
\usepackage[latin1]{inputenc}
\usepackage{color}
\usepackage[colorlinks]{hyperref}

\makeatletter
\def\blfootnote{\xdef\@thefnmark{}\@footnotetext}
\makeatother

\def\pa{\partial}
\def\vf{\varphi}

\def\ga{\gamma}

\def\dl{\delta}

\def\th{\theta}

\def\nn{\nonumber}

\def\wt{\widetilde}
\def\l{\left}
\def\r{\right}

\def\R{{\cal R}}

\def\b{{\sqrt b}}

\begin{document}

\title{\Large\bf Functions and Relations for an Evolving Star\\ with Spherical Symmetry}
\author{Ying-Qiu Gu}
\email{yqgu@fudan.edu.cn} \affiliation{School of Mathematical
Science, Fudan University, Shanghai 200433, China} \pacs{04.20.Dw,
04.70.-s, 97.60.Lf, 98.35.Jk}
\date{19th May 2017}

\begin{abstract}
In this paper, we drive and simplify some important equations and
relations for  an evolving star with spherical symmetry, and then
give some simple analysis for their properties and implications. In
the light-cone coordinate system, these equations and relations have
a normal and neat form which is much accessible than the usual
Einstein field equation. So they may be helpful for students to
study general relativity and for researchers to do further
discussion.

 \vskip 3mm \noindent{\bf Keywords:} {\sl stellar evolution,
field equation, singularity}

\end{abstract}
\maketitle

\section{Introduction}
\setcounter{equation}{0}

In general relativity, we have some typical exact solutions such as
the Schwarzschild, Curzon and Kerr metrics and some of their
extensions to the electrovacuum solutions such as the
Reissner-Nordstr\"om and Kerr-Newman metrics
\cite{exact1,exact2,exact3,exact6,exact4,gu3}. These solutions shed
some lights on the nature of spacetime and stars.

For a static and  spherically symmetrical space-time with perfect
fluid source, the line element is given by
\begin{eqnarray}
ds^2=B(r) d t^{~2} -A(r) dr^2-r^2(d\th^2+\sin^2\th
d\vf^2),\label{s1.1}
\end{eqnarray}
we have a number of exact solutions\cite{star1,star2,star3} and
generating method\cite{star4}. In these solutions  mass density
$\rho$ and pressure $P$ are usually expressed as the functions of
$(A,B)$ and their derivatives. In such expressions the properties of
the EOS is quite ambiguous, and then most of the solutions are
unrealistic in physics.

In fact, the  static asymptotically flat space-time with spherical
symmetry can be solved by the following procedure. The dynamics for
the space-time with perfect fluid can be reduced to the following
initial problem of an ordinary differential equation
system\cite{gu1,Sgrv},
\begin{eqnarray}
M'(r) &=& 4\pi G\rho r^2, \quad \qquad\qquad\qquad \qquad M(0)= 0,\label{s1.2} \\
\rho'(r) &=& -\frac {(\rho+P)(4\pi G P r^3+M)}{C_s^2 (r-2M)r},
\qquad \rho(0)=\rho_0, \label{s1.3}
\end{eqnarray}
in which $P=P(\rho)$ is the EOS of the fluid, $M(r)$ is the total
mass within the ball of radial coordinate $r$, $C_s=\sqrt{P'(\rho)}$
is the velocity of sound in the fluid. For any given $\rho_0>0$ we
get a unique solution. The metric components are given by
\begin{eqnarray}
A=\l(1-\frac {2M} r\r)^{-1},\qquad B=\exp\l(-\int^R_r \frac {2(4\pi
G P r^3+M)}{r(r-2M)}dr\r), \label{s1.4}
\end{eqnarray}
where $R<\infty$ is the radius of the star. $\rho(r)=P(r)=0$ in the
region $r\ge R$.

For any suitable EOS $P=P(\rho)$, the solution of (\ref{s1.2}) and
(\ref{s1.3}) can be easily solved numerically. By practical
calculation\cite{gu1}, we find that if the EOS satisfying the
following increasing and causal conditions
\begin{eqnarray}
0< C_s \le \frac 1 3, \qquad
P\to\l\{\begin{array}{lll} P_0\rho^\ga, & (\ga>1, \rho\to 0),\\
\frac 1 3 \rho,&(\rho\to \infty),\end{array}\r. \label{s1.5}
\end{eqnarray}
then all solutions of (\ref{s1.2}) and (\ref{s1.3}) are
singularity-free. That is, we always have $0\le\rho\le \rho_0$ and
$R_s=2M(R) < R$. The condition $\ga>1$ is necessary, which is caused
by  inertia of particles and  leads to the finite radius of the star
$R<\infty$ due to $C_s\to 0$ in (\ref{s1.3}).

However, for an evolutional star, a complete dynamical analysis
includes the hydrodynamics of matter, which is too complicated to be
solved. In this paper, we consider the simplest case, that is, a
star evolves with spherical symmetry. The results may be helpful to
understand the nature of a star and to do further researches.

\section{Dynamics for an evolving Star}
\setcounter{equation}{0}

The line element in the space-time generated by an evolving star
with spherical symmetry is generally given by\cite{Sgrv}
\begin{eqnarray}
ds^2=u^2 d\wt t^{~2} -(v d\wt t -w dr)^2-r^2(d\th^2+\sin^2\th
d\vf^2).\label{B1.1}
\end{eqnarray}
Here we take the light velocity $c=1$ as unit of speed. For a normal
star, $(u,v,w)$ are continuous functions of $(t,r)$ with suitable
smoothness. The null geodesic along the radius is described by
\begin{eqnarray}
(u+v)d\wt t - w dr=0. \label{B1.2.0}
\end{eqnarray}
Assume the solution is $f(\wt t,r)=C$, where  $C$ is a constant.
Making light-cone coordinate transformation $t =T(f(\wt t ,r))$,
where $T(f)$ is any smooth function satisfying $\pa_f T\pa_{\wt
t}f>0$, then we get the line element equivalent to (\ref{B1.1}) as
follows\cite{gu2,gu3},
\begin{eqnarray}
ds^2=a b dt^{2} + 2\b dt dr -r^2(d\th^2+\sin^2\th
d\vf^2),\label{B1.3}
\end{eqnarray}
where $(a,b)$ are continuous functions of $(t,r)$ with suitable
smoothness until the star becomes singular. (\ref{B1.3}) is similar
to the `Eddington - Finkelstein coordinates'. In this coordinate
system, the field equations have very simple form, and some of them
are integrable. However, the time coordinate $t$ is different from
the usual definition, which should be kept in mind. The usual
definition is given by $\dl\tau$,
\begin{eqnarray}
\dl\tau =\l(\sqrt{ab}dt+\frac 1 {\sqrt a} dr\r)c^{-2},\label{dft}
\end{eqnarray}
because in this time we have standard form
\begin{eqnarray}
ds^2=c^2\dl\tau^{2} -a^{-1} dr^2 -r^2(d\th^2+\sin^2\th
d\vf^2),\label{ds1}
\end{eqnarray}

For the external Schwarzschild solution, in coordinate system
(\ref{B1.3}), we have solution\cite{gu3}
\begin{eqnarray}
b=1,\qquad  a =1-\frac {R_s} r,\qquad ({\rm for}~ r\ge R
>R_s),\label{B1.4}
\end{eqnarray}
where $R=R(t)$ and $R_s$ are respectively the stellar radius and
Schwarzschild radius
\begin{eqnarray}
R_s\equiv 8\pi G\int^{R(t)}_0 \rho_{\rm grav} r^2 dr, \label{B1.25}
\end{eqnarray}
in which $\rho_{\rm grav}$ is the total gravitational mass-energy
density including influences of pressure and momentum. The
definition of $\rho_{\rm grav}$ is given below, and we find $R_s$ is
a constant.

Denote the 4-vector speed of the fluid by
\begin{eqnarray} U^\mu=\{U,~V,~0,~0\},\qquad
U_\mu=\{a b  U+\sqrt{b} V,~\b U,~0,~0\},\label{B1.5}
\end{eqnarray}
which satisfies the line element equation
\begin{eqnarray} 1=g_{\mu\nu} U^\mu
U^\nu=\l(a b U+2\b V\r)U.\label{B1.6}
\end{eqnarray}

For the perfect fluid model, the nonzero components of the
energy-momentum tensor $T_{\mu\nu}=(\rho+P)U_\mu U_\nu -Pg_{\mu\nu}$
are given by
\begin{eqnarray}
T_{tt}&=& b(\rho+P)\l(a\b  U+V\r)^2-a b  P,\label{B1.8}\\
T_{tr}&=& b(\rho+P)\l(a\b  U+V\r)U-\b P=T_{rt},\label{B1.9}\\
T_{rr}&=& b(\rho+P)~U^2,\qquad T_{\th\th}=Pr^2,\qquad
T_{\vf\vf}=Pr^2\sin^2\th,\label{B1.10}
\end{eqnarray}
where $P=P(\rho)$ is a given EOS which should satisfy  increasing
and causal conditions (\ref{s1.5}).

The nonzero components of Einstein tensor are given by
\begin{eqnarray}
G_{tt}&=&-\frac 1 r\l(\b\pa_t  a -a b\pa_r  a \r)-\frac 1 {r^2} a b  (1- a ),\label{B1.11}\\
G_{tr}&=&\frac 1 r \b\pa_r  a -\frac 1 {r^2} \b (1- a
)=G_{rt},\qquad G_{rr}= -\frac {\pa_r b
}{ra},\label{B1.12}\\
G_{\th\th}&=&\l(\frac  a  r \l( \frac {\pa_r b}{2b}+\frac {\pa_r
 a }{ a }\r)-\frac {1- a }{r^2}+\frac {\R} 2 \r)r^2,\qquad
G_{\vf\vf}=G_{\th\th}\sin^2\th,\label{B1.13}
\end{eqnarray}
where the scalar curvature  $\R$ depends on the second order
derivatives of the metric functions $(b, a )$. But it is not used in
the following discussion, because the related equations are not
independent, which can be derived from other equations.

By detailed calculations, we find only the following 3 equations are
independent ones in the Einstein equation $G_{\mu\nu}=-8\pi G
T_{\mu\nu}$,
\begin{eqnarray}
\pa_r b &=& 8\pi G r(\rho+P) b^2 U^2,\label{B1.14}\\
\pa_t  a  &=& 8 \pi G (\rho+P)r V\sqrt{b( a +V^2)}, \label{B1.15} \\
\pa_r  a  &=& -4\pi G r \l ((\rho-P)+(\rho+P)abU^2\r)+\frac {1- a }
r.\label{B1.16}
\end{eqnarray}
By  (\ref{B1.4}), (\ref{B1.25}) and (\ref{B1.15}), we learn $\pa_t
a=0$ if $r>R$, so $R_s$ is conserved for an evolving star.

Among the energy-momentum conservation law $T^{\mu\nu}_{~;\nu}=0$,
only the continuity equation $U_\mu T^{\mu\nu}_{~;\nu}=0$ is
independent, so we get
\begin{eqnarray}
U^\mu \pa_\mu \rho +(\rho+P)U^\mu_{~;\mu}=0.\label{B1.17}
\end{eqnarray}
Equations (\ref{B1.14})-(\ref{B1.17}) combined with EOS $P=P(\rho)$
form a closed system. The boundary conditions for  asymptotic flat
space-time are given by
\begin{eqnarray}
&{\rm at }~ &r=0:~~  a =1,\qquad \pa_r  a =0, \qquad~ V=0; \label{B1.23}\\
&{\rm at }~ &r=R:~  b=1,\qquad  a =1-\frac {R_s} {R}, \qquad \rho=0.
\label{B1.24}
\end{eqnarray}
Together with initial values $\{\rho(0,r), V(0,r)\}$,
(\ref{B1.14})-(\ref{B1.24}) has a unique solution.

\section{Simplification of the Equations}
\setcounter{equation}{0}

The equations derived above have a weakness, that is, the
geometrical variables $(a, b)$ and mechanical variables $(\rho, V)$
couple each other in a complicated manner, which increases the
difficulties for discussion. Besides, the physical meaning of
$(U,V)$ is unclear, which is quite different from the usual
definition $\frac {dr} {dt}$.

To simplify the relations, we introduce the following transformation
\begin{eqnarray}
U = \frac {1-v}{\sqrt{a b(1-v^2)}},\qquad V = \frac{\sqrt a~
v}{\sqrt{1-v^2}},\label{UV}
\end{eqnarray}
where the speed $|v|<1$ is approximately the usual definition.
Define an auxiliary energy function by
\begin{eqnarray}
F\equiv (\rho+P)abU^2=(\rho+P)\frac {(1-v)^2} {1-v^2}. \label{B2.1}
\end{eqnarray}
For a static star, we have $F=\rho+P$. Substituting  (\ref{UV}) and
(\ref{B2.1}) into (\ref{B1.14})-(\ref{B1.16}), we get simplified
relations
\begin{eqnarray}
\pa_r  a  &=& -8\pi G r \frac{\rho-P v}{1+v}+\frac{1-a}
r.\label{ar}\\
\pa_r b &=& 8\pi G r(\rho+P) \frac {b(1-v)}{a(1+v)},\label{br}\\
\pa_t  a  &=& 8 \pi G r(\rho+P) \frac{a \sqrt{b}~ v} {1-v^2},
\label{at}
\end{eqnarray}
Obviously, the geometrical variables $(a,b)$ are separated from
mechanical ones $(\rho, P, v)$. The solutions can be formally
expressed by
\begin{eqnarray}
a &=& 1-\frac {8\pi G} r \int^r_0\frac {\rho-Pv}{1+v} r^2 dr ,\label{sla}\\
b  &=& \exp\l(- 8 \pi G \int^R_r (\rho+P) \frac{(1-v)r}{(1+v)a}dr
\r), \label{slb}
\end{eqnarray}
and
\begin{eqnarray}
a = a(0,r)\exp\l( {8\pi G}\int^t_0 (\rho+P) \frac{\sqrt{b}~vr}
{1-v^2} dt\r),\label{slat}
\end{eqnarray}
By (\ref{sla}) and (\ref{slb}), for  an evolving star, we have
\begin{eqnarray} \rho_{\rm grav}=\frac 1 2
(\rho -P+F)=\frac {\rho-Pv}{1+v}. \label{B2.5}
\end{eqnarray}
For any $\rho(.,r)\in L^\infty([0,\infty))$, we have $a (\cdot,r)\in
C^0([0,\infty))$, and it has a positive minimum $ a _{\rm min}>0$.
$b(\cdot, r) \in C^1([0,R])$ is a monotonic increasing function of
$r$. For a normal star, the variables have the following range of
value,
\begin{eqnarray}
0<b\le 1,\qquad 0< a  \le 1,\qquad 0 \le \rho< \infty. \label{B1.22}
\end{eqnarray}

Simplifying (\ref{B1.17}) and the consistent equation of
(\ref{B1.15}) and (\ref{B1.16}) $\pa_{tr} a = \pa_{rt}a$, we get the
dynamical equation for $(\rho,v)$, which is a first order hyperbolic
differential equation system,
\begin{eqnarray}
(1-C_s^2)\frac {\pa_t \rho }{a\sqrt b}&+&(v+C_s^2)\frac{\pa_r \rho} {1-v} +(\rho+P) \frac{\pa_r v}{(1-v)^2}  \nn\\
&=& \frac{(\rho+P)}{a(1-v)}\l( 4\pi G r(\rho v-P) -\frac 1 {2r} [1-a+(1+3a)v]\r)  , \label{B1.19}\\
(1-C_s^2) \frac {\pa_t v}{a\sqrt b}& +&(1+v)^2 C_s^2 \frac{\pa_r \rho}{\rho+P}+ (v+C_s^2) \frac {\pa_r v}{1-v}  \nn\\
&=& \frac {1+v} a\l( 4\pi G r(C_s^2\rho v-P) -\frac 1 {2r} [1-a
+(1+3a)C_s^2 v]\r). \label{B1.20}
\end{eqnarray}
The characteristic speeds are given by
\begin{eqnarray}
V_1 = \frac { a\sqrt b(v+C_s)}{(1-C_s)(1-v)},\qquad V_2 = \frac {
a\sqrt b (v-C_s)}{(1+C_s)(1-v)}.\label{crs}
\end{eqnarray}
The disturbance of the solution $(\rho,v)$ propagates at such speed
$\frac {dr}{dt}=V_k$.

Substituting (\ref{sla}) and (\ref{slb}) into (\ref{B1.19}) and
(\ref{B1.20}), we get closed equations for $(\rho,v)$, which include
all information for an evolving star. Combining (\ref{B1.19}) and
(\ref{B1.20}) with initial and boundary conditions, we have a unique
solution. The numerical solution can be easily solved by method of
characteristics.

\section{Analysis for the solutions}
\setcounter{equation}{0}

Since the dynamical equations (\ref{B1.19}) and (\ref{B1.20}) are
quite complicated, and the rigorous solution is absent. So we can
only qualitatively analyze some asymptotic properties of the
solutions, and shed some lights on the behavior of an evolving star.

Obviously, when $\rho<\infty$, the star is normal and the spacetime
should be singularity-free. $\rho\to\infty$ is a necessary condition
for the spacetime becoming singular, and $a\to 0$ or $b\to 0$ is the
signal that the spacetime becomes singular.

In (\ref{slat}), $a (0,r)$ is determined by the initial distribution
$\rho(0,r)_{\rm grav}$ via (\ref{sla}), so we have $a>0$. If
$\rho\to\infty$, we have $P\to C^2_0 \rho$. So according to
(\ref{slb}), $b(t,0)\to +0$ only if $\rho(t,r)\ge\rho_0(t)
r^{-2},(r\to 0)$. This can be checked as follows. For the critical
initial distribution
\begin{eqnarray}
\rho\to \frac {\rho_0}{r^2},\qquad P\to  C_0^2 \rho\to \frac {\rho_0
C_0^2}{r^2}, ~(r\to 0).\label{B3.15}
\end{eqnarray}
When $r\to +0$, we have $( a \to 1,~v\to 0)$.  For any given $r_0$
satisfying $0<r_0-r\ll R$, by (\ref{slb}), we have estimation
\begin{eqnarray}
b(t,r) &= &\exp\l(-8\pi G (\int^R_{r_0} +\int^{r_0}_r )(\rho+P) \frac{(1-v)r}{(1+v)a} dr\r)\nn\\
&\to& A_1(t,r_0) \exp\l(-8\pi G\l(1+C_0^2\r) \int^{r_0}_r \frac {
\rho_0}{r} dr
\r)\nn\\
&=& A_2(t,r_0) r^{8\pi G \rho_0 \l(1+C_0^2\r)}\to 0,
~~(r\to0),\label{4.16}\end{eqnarray} in which all $0<A_k<\infty$ are
independent of $r$. (\ref{4.16}) means the space-time itself becomes
singular. However, for $\rho(t,r)\to \rho_0 (t)r^{-n},(0<n<2)$, we
still have $b>0$, and the space-time is still measurable, although
the curvature becomes singular.

In the case of a star with extreme high temperature and pressure,
the EOS of the fluid becomes simple. We have the following
approximations,
\begin{eqnarray}
P\dot=C_0^2 \rho,\qquad C_0\le \frac {\sqrt{3}}3. \label{B2.13}
\end{eqnarray}
Noticing $v(t,0)=0$, by (\ref{sla}), (\ref{slb}) and (\ref{slat}),
near the center $r=0$, we have
\begin{eqnarray}
|v| \ll 1,\qquad b \dot= b_0(t),\qquad  a \dot= 1.\label{B2.15}
\end{eqnarray}
Substituting (\ref{B2.13}), (\ref{B2.15}) into (\ref{B1.19}) and
(\ref{B1.20}), we have the simplified dynamical equations which hold
near the center of  a star,
\begin{eqnarray}
\pa_r w &=& \pa_\eta w+ w \l(1+C_0^{-2}\r)\pa_\eta v +4\pi
G\l(1+C_0^2\r)r ,\label{eqw} \\
\pa_r v &=& \pa_\eta v+w^{-1}\l(1+C_0^2\r)^{-1}\pa_\eta w -\frac{2v}
r, \label{eqv}
\end{eqnarray}
where $d\eta\equiv\sqrt{b_0} dt$  and $w\equiv \rho^{-1}$. Since we
have not made approximation for $\rho$, so (\ref{eqw}) and
(\ref{eqv}) are valid for all range value of $\rho$.

(\ref{eqw}) and (\ref{eqv}) has a static solution
\begin{eqnarray}
v=0,\qquad \rho=\frac 1 w=\frac 1 {C+2\pi G\l(1+C_0^2\r)r^2},
\label{slp}
\end{eqnarray} where $C\ge0$ is a constant.
Even for (\ref{eqw}) and (\ref{eqv}) with constant characteristic
speeds, we can hardly find a rigorous evolving solution. However,
(\ref{eqw}) is integrable under some ansatz. If we set $v=f(t+r)r$,
in this case we have
\begin{eqnarray}
\frac 1 \rho = w = \l( \frac 1{\rho_0(\eta+r)}+\frac{4\pi G
C_0^2}{f'(\eta+r)} \r) \exp \l(-\frac {1+C_0^2} {2C_0^2}
\eta(\eta+2r) f'(\eta+r)\r)- \frac{4\pi G C_0^2}{f'(\eta+r)}.
\label{slw}
\end{eqnarray}
Of course, under such ansatz, (\ref{eqv}) is not strictly satisfied
in general. For reasonable function $f(t+r)$, (\ref{slw}) can
display local evolving trend of mass density $\rho$ as $(r\to
0,\eta\to 0)$.

\section{Discussion and conclusion}
\setcounter{equation}{0}

\begin{enumerate}

\item
We derived the dynamics for an evolving star in the light-cone
coordinate system.  Under some transformation of variables, the
equations and relations of dynamics take simple and neat form, which
is much more accessible than the usual field equations. The final
dynamical equation is reduced to a standard first order hyperbolic
system (\ref{B1.19}) and (\ref{B1.20}), which can be discussed by
method of characteristics.

\item
The singularity analysis for an evolving star is equivalent to
discuss the singularity of equation (\ref{eqw}) and (\ref{eqv}) near
the center $r\to+0$. This is almost a linear first order hyperbolic
system with constant characteristic speed.

\item
The analysis shows that, it is the EOS of matter and the initial
distribution of mass density and speed  rather than the total
mass-energy to decide the fate of an evolving star.
\end{enumerate}

\section*{Acknowledgments}
The author is grateful to his supervisor Prof. Ta-Tsien Li and
Prof. Han-Ji Shang for their encouragement.

\end{document}